\documentclass[sigconf]{acmart}
\renewcommand\footnotetextcopyrightpermission[1]{} 
\usepackage{amsmath}
\usepackage{placeins}
\usepackage{graphicx}
 \usepackage{caption}
 \usepackage{subcaption}

\AtBeginDocument{%
  \providecommand\BibTeX{{%
    \normalfont B\kern-0.5em{\scshape i\kern-0.25em b}\kern-0.8em\TeX}}}
\acmConference[KDD '21]{}{August 14--18, 2021}{Virtual Conference}
\begin{document}
\title{Retinal Microvasculature as Biomarker for Diabetes and Cardiovascular Diseases}
\author{Anusua Trivedi}
\authornote{Authors contributed equally to this research.}
\affiliation{\institution{Microsoft}}
\email{antriv@microsoft.com}

\author{Jocelyn Desbiens}
\authornotemark[1]
\affiliation{\institution{Intelligent Retinal Imaging Systems}}
\email{jdesbiens@retinalscreenings.com}

\author{Ron Gross, MD}
\authornotemark[1]
\affiliation{\institution{Intelligent Retinal Imaging Systems}}
\email{rgross@retinalscreenings.com}

\author{S. Gupta, MD}
\affiliation{\institution{Retina Specialty Institute}}
\email{sgupta@retinalscreenings.com}

\author{Rahul Dodhia}
\affiliation{\institution{Microsoft}}
\email{radodhia@microsoft.com}

\author{Juan Lavista Ferres}
\affiliation{\institution{Microsoft}}
\email{jlavista@microsoft.com}
\renewcommand{\shortauthors}{Trivedi et al.}
\begin{abstract}
\noindent\textbf{Purpose}: To demonstrate that retinal microvasculature per se is a reliable biomarker for Diabetic Retinopathy (DR) and, by extension, cardiovascular diseases.\newline \textbf{Methods}: Deep  Learning  Convolutional  Neural  Networks  (CNN)  applied to color fundus images for semantic segmentation of the blood vessels and severity classification on both vascular and full images. Vessel reconstruction through harmonic descriptors is also used as a smoothing and de-noising tool. The mathematical background of the theory is also outlined.\newline\textbf{Results}: For diabetic patients, at least 93.8\%  of  DR  No-Refer \textit{vs.} Refer classification can be related to vasculature defects. As for the Non-Sight Threatening \textit{vs.} Sight Threatening case, the ratio is as high as 96.7\%.\newline\textbf{Conclusion}: In the case of DR, most of the disease biomarkers are related topologically to the vasculature.\newline\textbf{Translational Relevance}: Experiments conducted on eye blood vasculature reconstruction as a biomarker shows a strong correlation between vasculature shape and later stages of DR.

\end{abstract}
\maketitle
\thispagestyle{empty}
\section{Introduction}
\label{sec:introduction}
Diseases that primarily affect blood vessels, including diabetes and cardiovascular disease, are four of the top seven causes of death in the United States and have an increasing prevalence worldwide \cite{nvs:2017}. An enhanced ability to diagnose these diseases and their stage and determine an individual's propensity to develop or have progressive complications from these diseases would be of enormous benefit. Recent technological advances have allowed the development of clinical data acquisition and analysis to begin to fulfill this goal. Retinal vasculature provides a unique ability to examine blood vessels and can now be imaged non-invasively with efficient and effective devices in large populations. The benefits of telemedicine using these images to diagnose and determine the severity of diabetic retinopathy have been demonstrated to be as accurate as live examinations, if not more \cite{lin:2002, man:2015}. The ability to incorporate this capability into a health system to improve patient outcomes has been shown in the English NHS Diabetic Eye Screening Program that has reported an 82.8\% rate of diabetic retinopathy evaluation, over time resulting in diabetic retinopathy no longer being the leading cause of blindness and visual disability among working-class adults in the UK \cite{scan:2017}.

Artificial Intelligence and Deep Learning provide new technology to facilitate the ability to improve the sensitivity and specificity of the analysis of retinal images. We can use a deep convolutional neural network-based model to detect referable DR \cite{say:2019}. It has been shown that Deep Learning assistance can improve the accuracy of reader grading of retinal images for DR severity \cite{gul:2016}. Additionally, to further improve access, quality, and cost-effectiveness, artificial intelligence has been FDA cleared for autonomous diagnosis of more than mild diabetic retinopathy and diabetic macular edema using a single camera in patients without diabetic retinopathy. The pivotal FDA trial demonstrated a sensitivity of 87.2\% and specificity of 90.7\% \cite{abr:2018}.

Fundus photography, with the assessment of retinal vascular measurements, is useful in the evaluation of other vital diseases in addition to diabetes, including cardiovascular, cerebrovascular, and Alzheimer's diseases. Retinal vascular caliber has been shown to correlate with diabetes complications: retinopathy, nephropathy, neuropathy, and cardiovascular risk \cite{ikr:2017, ho:2017}. The impact of including retinal parameters and biomarkers with traditional cardiovascular risk measures in a diabetic population showed a statistically significant improvement based on net re-classification improvement \cite{ho:2017}. This improved ability to assess cardiovascular risk has been demonstrated in an older population in \cite{won:2017}. Given the similar embryological origins and structural characteristics of retinal and cerebrovascular vessels, it is not surprising to find associations between retinal signs and brain microvascular disease \cite{han:2014}. There is also evidence of an increased abnormality of several retinal vascular parameters in patients with Alzheimer's Disease compared to matched controls \cite{che:2014}.

\begin{table*}[!h]
\centering
\renewcommand{\arraystretch}{1.5}
\begin{tabular}{c|c|c|c|c|c}
&\textbf{Normal}&\textbf{Mild}&\textbf{Moderate}&\textbf{Severe}&\textbf{Proliferative}\\
\hline
\textbf{Normal}&\textbf{1436}&55&6&0&0\\
\textbf{Mild}&14&\textbf{448}&260&2&0\\
\textbf{Moderate}&8&12&\textbf{1652}&252&56\\
\textbf{Severe}&0&2&19&\textbf{1286}&179\\
\textbf{Proliferative}&4&1&7&30&\textbf{1259}\\
\end{tabular}
\vspace{0.3cm}
\caption{\small Confusion matrix of clinician interpretation (clinician-1 row-wise and clinician-2 column-wise) in terms of DR severity at eye level}
\label{table:kappa}
\end{table*}

In medical images, identifying candidate regions is of the highest importance since it provides intuitive illustrations for doctors and patients of how the diagnosis is inferred. Recently, advances in Deep Learning have dramatically improved the performance of disease detection. Most of these Deep Learning systems treat CNN as a kind of black box, lacking comprehensive explanation. Traditional ways of visual "feature finding" are given through heat-map generation or sliding windows. In this report, we show that for retinal diseases, most of the visual features allowing AI recognition like vessel curvature, angles, \textit{etc}. are topologically concentrated along the retinal vasculature.

\section{Prior Work}
Biomarkers are traditionally associated with lesions on the retina caused by the disease, such as Microaneurisms, Hemorrhages, Exudates, Macular Edema, and Neovascularization \cite{jen:2015}. Moreover, they are also related to typical risk factors for DR (HbA1c, blood pressure, total cholesterol, etc.). The absence or presence, type, and severity of retinal vessel lesions diagnosed by retinal photography are biomarkers of DR status and are commonly used in screening clinics and research \cite{tho:2015}. Many methods from Image Processing and Deep Learning domains have been designed and implemented to automatically or semi-automatically detect the markers.

Retinal markers like the retinal arteriolar and venular caliber and their arterio-venous ratio (AVR) are of great importance. It has been found useful for early diagnosis of diseases such as hypertension, diabetes, stroke, other cardiovascular diseases in adults, and retinopathy of premature \cite{ikr:2004}. One calculates it as the ratio of average arteriolar diameter and average venous diameter of the vessels within one and two disk diameter circular rings from the optic disk center \cite{knu:2003}. So only a small but essential part of the vasculature is used.

Retinal vessel geometry comprises another group of retinal vessel-based bio\-markers. It includes measures of vessel branching angles, branching complexity, tortuosity, length-to-diameter ratio, and fractals \cite{lim:2017, che:2012}. They have been used, among other things, to identify patterns summarizing the retinal vascular network in the elderly and to relate them to cardiovascular history (see Table 2 of \cite{arn:2018} for a full list of geometric biomarkers).

The extraction of retinal microvasculature with the help of Deep Learning, which dates back to 2015 \cite{soo:2019}, has shown outstanding results and outperformed drastically the methods based on traditional image processing \cite{alm:2018}. This allows the segmentation of vessels to the extent of being able to characterize the whole vasculature as a biomarker encompassing, therefore, other biomarkers such as tortuosity, curvature, angles, \textit{etc}.

\section{Dataset}
This study is performed utilizing the Intelligent Retinal Imaging Systems (IRIS, Pensacola, Florida) FDA-cleared, HIPPA compliant, commercial platform data\-base. The IRIS software is a software-as-a-service (SaaS) application that is hosted on the Internet which gives clinicians the ability to scan a patient's retina with a fundus camera, store the images in a cloud-based data-store, view the images and their associated information, and offer an opinion on the scans. 

The dataset consisted of color digital fundus photography obtained in a non-Eye-Care Professional's environment operated by staff with minimal training. The IRIS software solution is designed to be camera agnostic. For this study, a total of 6,988 graded $45^\circ\times40^\circ$ CenterVue DRS camera (CenterVue SpA, Freemont, CA) were utilized. All images are monoscopic, and single images centered on the fovea. These 6,988 graded images represent a random sample from over 430,000 orders within the IRIS database.

The categorization of the images was performed using the Diabetic Retinopathy Severity Scale (DRSS) based on the International Classification of Diabetic Retinopathy (ICDR) criteria. Ground truth for each image was determined by the agreement of two experienced grading board-certified ophthalmologists. If there was disagreement among the two, a third ophthalmologist acted as an adjudicator. All images were evaluated in the standard color image as well as an IRIS proprietary image enhancement. We see from the confusion matrix presented in Table \ref{table:kappa} that the main disagreements lie along the Mild/Moderate boundary. 

Agreements between the clinicians were assessed by Cohen's $\kappa$ coefficient, whose equation is $\kappa=\frac{p-q}{1-q}$, where $p$ is the observed probability of agreement, and $q$ is the probability of chance agreement. The golden rule being the more significant the $\kappa$, the better is the agreement. The $\kappa$ value of the confusion matrix is $0.838$, which shows a pretty good agreement. 

\begin{figure*}[t]
	\centering
	\includegraphics[width=0.7\textwidth]{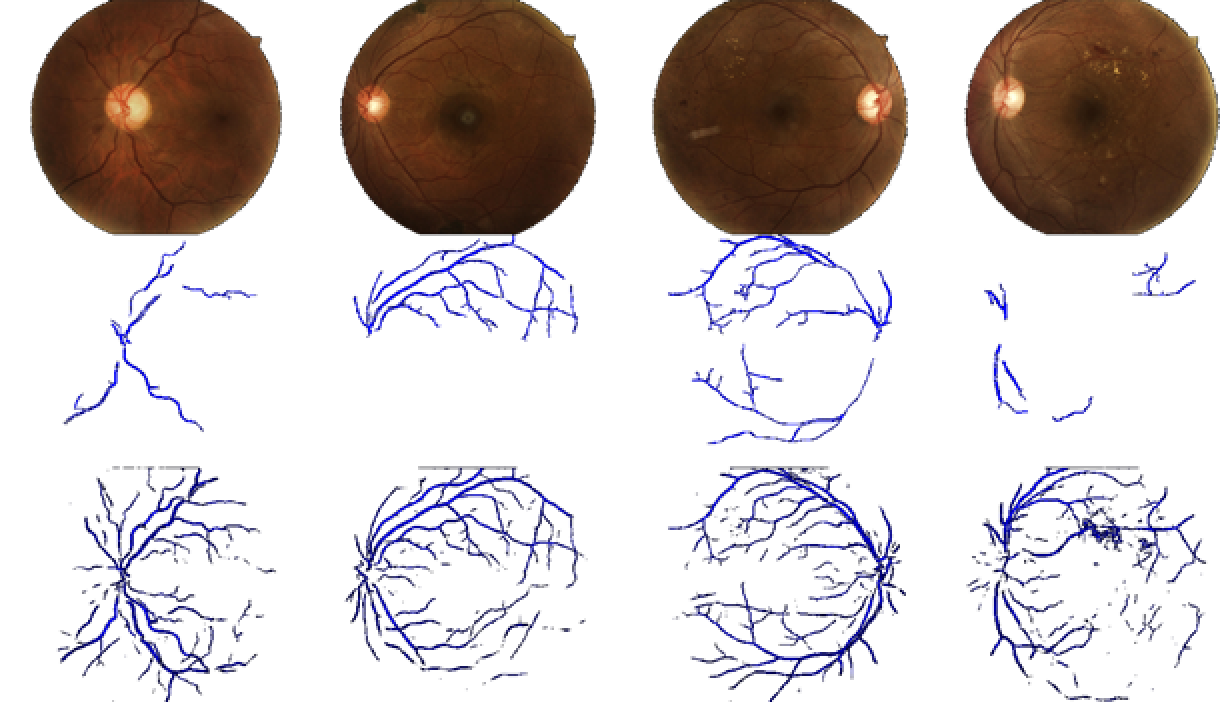}%
	\caption{\small Image processing (second row) and semantic segmentation (third row) of the same color fundus images (first row)}
	\label{fig:vasc1}
\end{figure*}

\begin{figure*}[tb]
	\centering
	\includegraphics[width=0.6\textwidth]{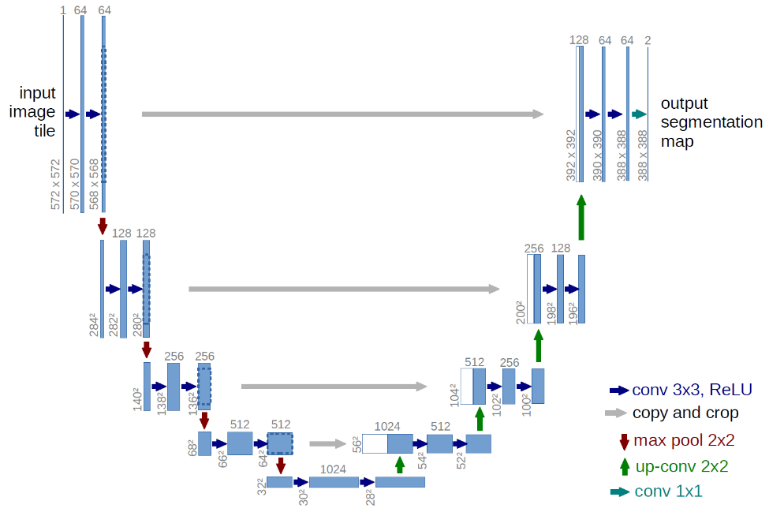}%
	\caption{\small Deep Learning model architecture used for semantic segmentation}
	\label{fig:vasc2}
\end{figure*}

\section{Methods}
Due to the time-consuming and repetitious aspects of manual blood vessel segmentation, automatic segmentation of retinal vessels is necessary for making a computer diagnostic system for ophthalmic disorders. Automatic segmentation of the blood vessels in retinal images is important in detecting several eye diseases because, in some cases, they affect the vessel network itself.  

\subsection{Image Processing Segmentation}
Several topological features of retinal vessels (\textit{e.g.} diameter, length, branching angle, tortuosity, curvature) have diagnostic value\cite{estrada:2015}. They can be used in the follow-up of disease progression, treatment, and evaluation of various cardiovascular and eye-related diseases (\textit{e.g.} diabetes, hypertension, arteriosclerosis, and neovascularization)\cite{macgillivray:2014}. The basic image segmentation is obtained by applying the following filters to the source image: 
\begin{itemize}
    \item Gray-scale conversion of the color fundus image. 
    \item Standardization. 
    \item Contrast-limited adaptive histogram equalization (CLAHE). 
    \item Gamma adjustment.
\end{itemize}

However, due to the high variance in image quality and camera type, image processing segmentation software tends to deliver good to poor accurate vascular segmentation. In Figure~\ref{fig:vasc1}, the middle row displays blood vessel networks segmented through image processing. As one can see, some images have almost no segmented vasculature. For around one-third of the dataset, images have partial or no segmentation due to image quality, intensity variation, lens artifacts, \textit{etc}. This is the reason why we need a more accurate segmentation tool.

\begin{figure*}[!t]
    \centering
    \begin{subfigure}{0.45\textwidth}
        \centering
        \includegraphics[scale=0.22]{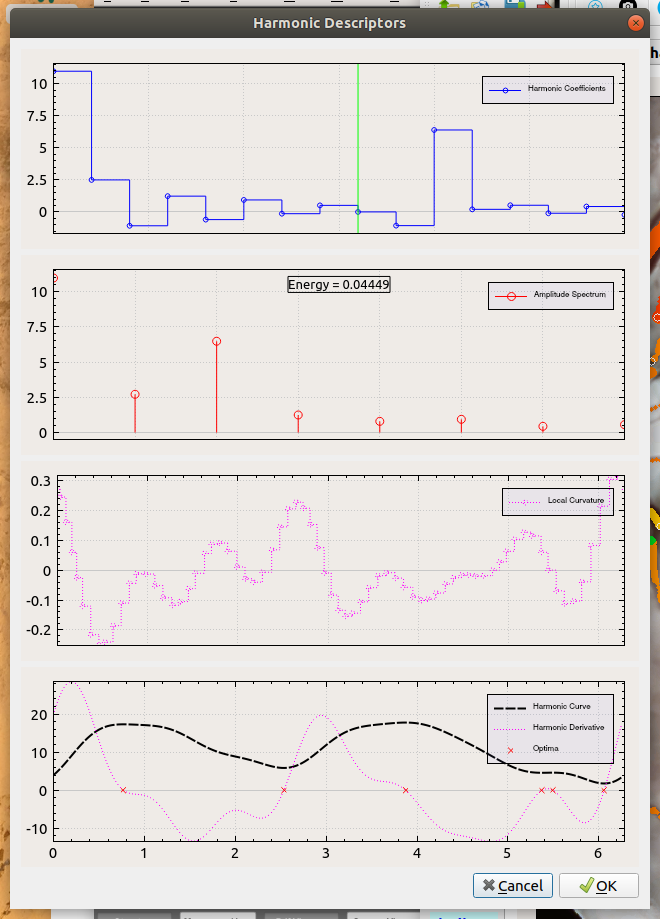}%
        \caption{\small Reconstructed vessel from harmonic descriptors. Bottom window contains the reconstructed vessel and its derivative. Local curvature is shown in third window from top.}
    \end{subfigure}%
    \begin{subfigure}{0.45\textwidth}
        \centering
        \vspace{1.9cm}
        \includegraphics[scale=0.12]{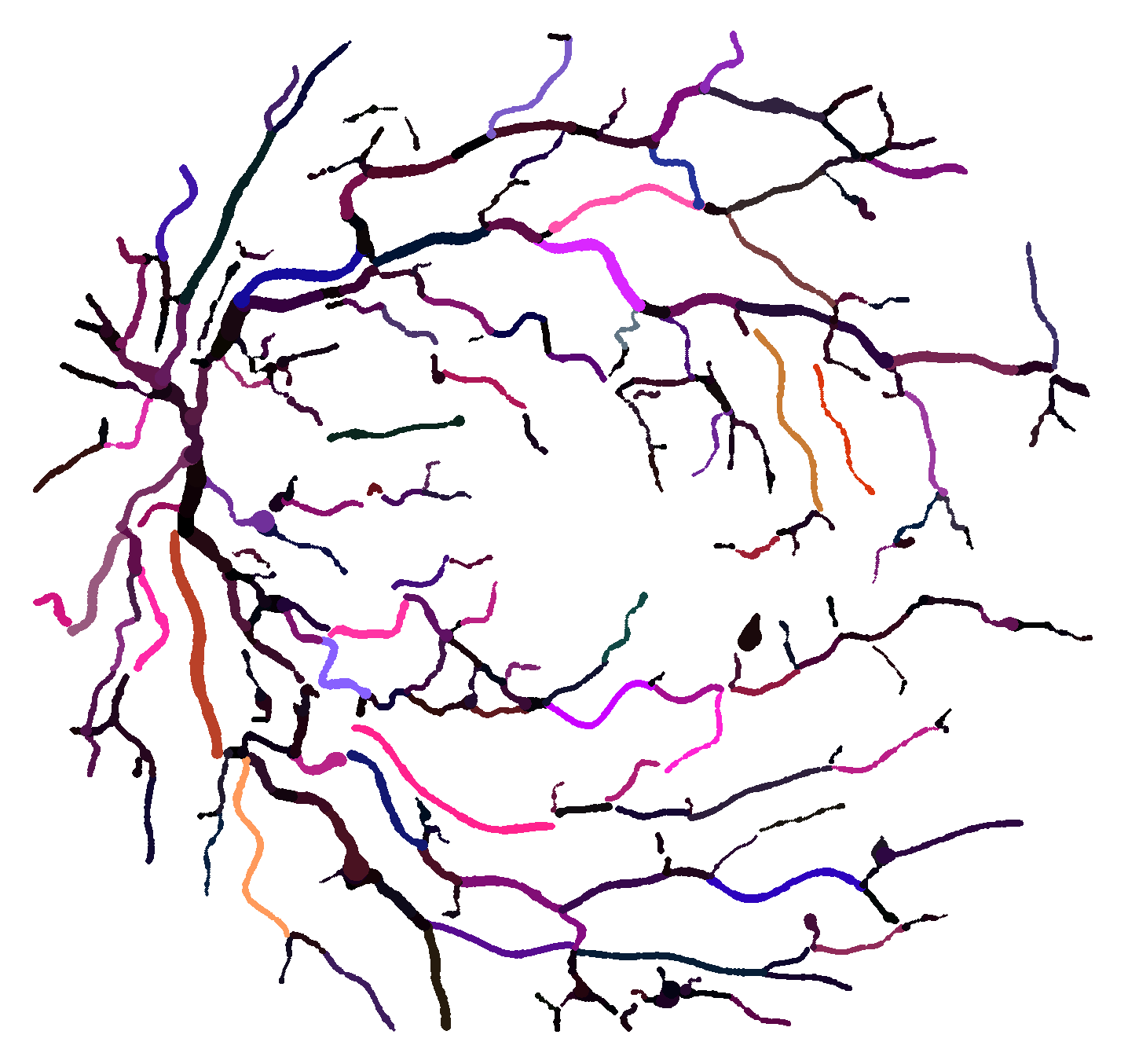}%
        \caption{\small Harmonic reconstruction of the vasculature of a fundus image. RGB colors are respectively assigned the scale invariant values of $|(a_0,b_0)|$, $|(a_2,b_2)|$, and $|(a_1,b_1)|$. Implementation relies heavily upon \texttt{FFTW} \cite{fri:1999}.}
    \end{subfigure}
    \caption{Harmonic descriptors and vessel reconstruction}
      \label{fig:harmonic-image}
\end{figure*}
\subsection{Semantic Segmentation}
In recent years, Deep Learning has achieved great success in visual perception, and the semantic segmentation of images is one of the most successful cases. The overall architecture of the model we used is shown in Figure~\ref{fig:vasc2}. Our model adopts a full convolutional neural network, which is commonly used in most semantic segmentation tasks. It can be briefly divided into two parts - encoder and decoder. The encoder is a convolutional neural network that extracts features from the input image, such as the retinal vasculature image. The decoder will up-sample the extracted features to the resulting image that we desired, such as the vessel segmentation in our case.

The neural network architecture is derived from the \texttt{U-Net} architecture \cite{olaf:2015}. The major advantage of this architecture is its ability to consider a wider context when predicting a pixel. This is facilitated by using the large number of channels used in the up-sampling operation. 

We apply cross-entropy for the loss function, and the stochastic gradient descent (\texttt{SGD}) is employed for back-propagation optimization. After each convolutional layer, the activation function is the Rectifier Linear Unit (\texttt{ReLU}). A dropout rate of 0.2 is used between two consecutive convolutional layers. Among the dataset, 150 images with correct image processed segmentation were chosen for training purposes. The training was performed over 50 epochs, with a mini-batch size of 16 patches. Performance obtained was measured by a 97.9\% AUC (see Figure~\ref{fig:semantic_roc}) on the 150 training images.

The experiments were run on a high-end GPU, the \texttt{Tesla V100-PCIE}, containing 2496 CUDA cores. After training, all 6,988 images in the dataset were processed to generate the vasculature images (see bottom row of Figure~\ref{fig:vasc1}). Segmentation time took 1 hour and 49 minutes.

\subsubsection{Vessel Reconstruction and Harmonic Descriptors}
The microvasculature segmentation one obtains from Deep Learning has better coverage than image processing can yield, but it is far from being perfect. Noise reduction has to be provided, blobs and false-positive pixels have to be removed from the processed images. Furthermore, some vessel smoothing has to be applied to the generated segments. We based our smoothing method on Fourier Descriptors \cite{zah:1972}. We give a mathematical description of the theory in Appendix~\ref{theory}.

\paragraph{Image Processing Steps}
 (\textbf{1}) Binarize vasculature image obtained by semantic segmentation.  (\textbf{2}) Extract skeleton from binary image by thinning. (\textbf{3}) Build an abstract graph from the skeleton. (\textbf{4}) Remove noise from graph and apply vessel tracking over the entire vasculature. (\textbf{5}) Reconstruct each extended vessel in the graph by harmonic descriptors smoothing.

\paragraph{Harmonic Descriptor Smoothing}
Harmonic descriptors have multiple advantages. For instance, knowing the harmonic descriptors of a given vessel, it is straightforward to compute its first and second derivatives, yielding the local curvature. A by-product, i.e. averaging the curvature over the entire vasculature, gives an estimation of the tortuosity of the vessel network. Likewise, the angle made by two incident vessels can be readily computed.

\begin{figure*}
	\centering
	\includegraphics[width=0.7\textwidth]{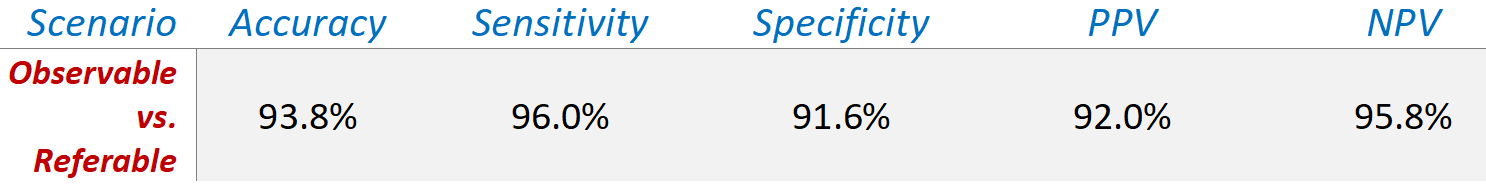}%
	\caption{\small Classification results (\textbf{I-A})}
	\label{fig:I_A}
\end{figure*}

\begin{figure*}
	\centering
	\includegraphics[width=0.7\textwidth]{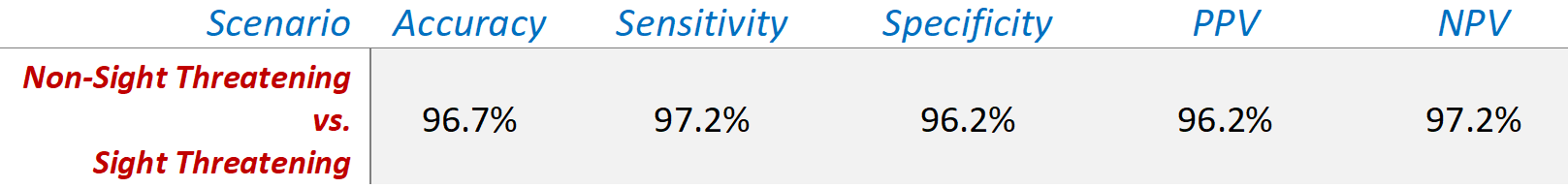}%
	\caption{\small Classification results (\textbf{I-B})}
	\label{fig:I_B}
\end{figure*}

\begin{figure*}
	\centering
	\includegraphics[width=0.7\textwidth]{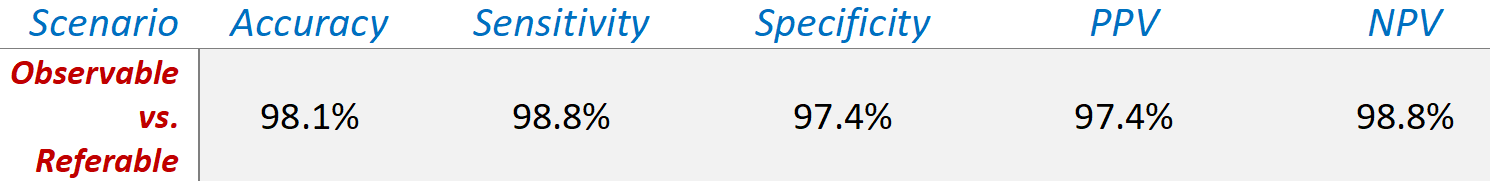}%
	\caption{\small Classification results (\textbf{II-A})}
	\label{fig:II_A}
\end{figure*}

\begin{figure*}[!h]
	\centering
	\includegraphics[width=0.7\textwidth]{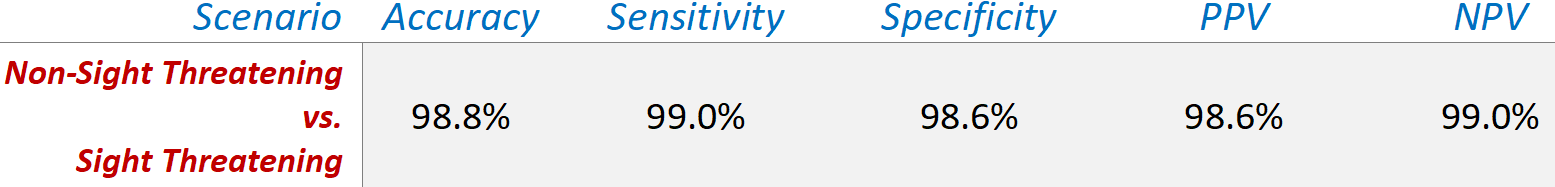}%
	\caption{\small Classification results (\textbf{II-B})}
	\label{fig:II_B}
\end{figure*}

\section{Experiments}
As retinal blood vessels are the only deep vessels that can be observed in the human body, they could directly reflect the state of some cardiovascular diseases. They could also reflect the severity of DR \cite{fraz:2012}. We ran four experiments to test this assumption.

\subsection{Severity Classification of Vasculature Images}
To obtain a baseline measure, we run experiments against the corresponding reconstructed vasculature images without the out-of-vessels lesions (Scenario I). A technique called fine-tuning is used to make a pre-trained model adapt to a new dataset. We first train a model on a large dataset with millions of images. Then on the second step, we adjust the weights using the domain-specific dataset. Fine-tuning often benefits medical domains with relatively small datasets. At the same time, fine-tuning carries the accuracy advantage of deep learning. In our case, we used the pre-trained \texttt{ResNet101} model \cite{he:2015}. The training was performed over 50 epochs, with a mini-batch size of 8 images. Learning rate per mini-batch was $=[0.01] * 5 + [0.001] * 5 + [0.0001] * 5 + [0.00001]$ with a $l^2$ regularization weight of $0.0005$. On a \texttt{Tesla V100-PCIE}, training time amounted to 56 minutes.

\subsubsection{Scenario I-A}
First dataset was Normal + Mild \textit{vs} Moderate + Severe + Proliferative (No-Refer \textit{vs} Refer) containing segmented retinal vasculature images only. Training dataset consists of 5,988 vasculature images, validation dataset of 1,000 vasculature images.  Classification results are to be found in Figures~\ref{fig:I_A} and~\ref{fig:I_ROCS}(a).

\subsubsection{Scenario I-B}
Second dataset was Normal + Mild \textit{vs} Severe + Proliferative (Non-Sight Threatening \textit{vs} Sight Threatening) containing segmented retinal vasculature images only. Training dataset consists of 4,008 vasculature images, validation dataset of 1,000 vasculature images. Classification results are to be found in Figures~\ref{fig:I_B} and~\ref{fig:I_ROCS}(b).

\subsection{Severity Classification of Full Retinal Images}
To check if lesions can discriminate more than vasculature alone, we re-run same experimentes on the full images containing all potential lesions (Scenario II). We used the same training/validation datasets as in the above experiments.

\subsubsection{Scenario II-A}
Third dataset was Normal + Mild \textit{vs} Moderate + Severe + Proliferative (No-Refer \textit{vs} Refer) containing full retinal images only. Training dataset consists of 4,008 full images, validation dataset of 1,000 full images.  Classification results are to be found in Figures~\ref{fig:II_A}, and ~\ref{fig:II_ROCS}(a).

\subsubsection{Scenario II-B}
Fourth dataset was Normal + Mild \textit{vs} Severe + Proliferative (Non-Sight Threatening \textit{vs} Sight Threatening) containing full retinal images only. Training dataset consists of 5,988 full images, validation dataset of 1,000 full images.  Classification results are to be found in Figures~\ref{fig:II_B}, and ~\ref{fig:II_ROCS}(b).

\begin{table*}[!h]
\centering
\renewcommand{\arraystretch}{1.5}
\resizebox{0.7\textwidth}{!}{
\begin{tabular}{c|c|c}
&\textbf{A}&\textbf{B}\\
\hline
\textbf{Vascular Images}&93.8\%&96.7\%\\
\hline
\textbf{Full Images}\text{ (Baseline)}&98.1\%&98.8\%\\
\hline
\textbf{Mean Vasculature Tortuosity}&\shortstack{\text{Observable - }2.86\\ \text{Referable - }3.14}&\shortstack{\text{Non-Sight Threatening - }2.86\\ \text{Sight Threatening - }3.26}\\
\end{tabular}
}
\vspace{0.3cm}
\caption{Summary of accuracy measures and mean tortuosity for both scenarios}
\label{table:scenarios}
\end{table*}

\section{Results}
For scenario \textbf{A}, we can say that 95.6\% $\left(=\frac{93.8\text{\%}}{98.1\text{\%}}\right)$ of the success rate can be explained by the blood vessel defects present in the microvasculature. In case of scenario \textbf{B}, 97.9\% $\left(=\frac{96.7\text{\%}}{98.8\text{\%}}\right)$ of the success rate can be attributed to the vasculature itself.


As everyone knows the low-level layers of a convolutional neural network match simple salient features such as lines (vertical, horizontal, and diagonal), corners, and contours \cite{zei:2013}. Other intermediate level layers might, for instance, match curved lines and circles. Actually, in CNNs, each layer of nodes is trained on a distinct set of features based on the previous layer's outcome. The deeper down into the neural network, the simpler the features the nodes can recognize, obviously because they aggregate and recombine features from the previous layer. As a matter of fact, related retinal vessel diseases tend to modify the morphological and topological structure of the vasculature by creating new vessels (neovascularization), increasing vessel curvature, widening branching angles, \textit{etc}. 

We make the hypothesis that the deep learning recognized geometric features are  essential to help CNNs provide such good accuracy results. As shown by Cheung CY et. al.~\cite{cheung2011retinal}, the retinal vascular tortuosity from retinal images may provide further information regarding effects of cardiovascular risk factors. From our experiments, the success rate for cardiovascular risk detection can be attributed to the success rate of vasculature itself due to the fact that tortuosity as overlay on the vasculature image brings a differential layer of information, with its delta being larger in \textbf{B} than in \textbf{A} (see \textbf{Mean Vasculature Tortuosity} row in  Table~\ref{table:scenarios}). 

\section{Conclusion}
There is a strong interest in the early detection of individuals with diabetes and hypertension. It is known that individuals with impaired glucose metabolism or diabetes have higher mortality from cardiovascular disease \cite{bru:2006}. In a similar fashion, individuals with high to standard blood pressure or pre-hypertension are more likely to develop cardiovascular events \cite{hsi:2007}. The new tortuosity definition we explain in this paper can provide a high level of abstraction similar to what CNNs are offering and as such can help us better identify cardiovascular risks in patients.

\bibliographystyle{ACM-Reference-Format}
\bibliography{sample-sigconf}


\begin{thebibliography}{32}


\ifx \showCODEN    \undefined \def \showCODEN     #1{\unskip}     \fi
\ifx \showDOI      \undefined \def \showDOI       #1{#1}\fi
\ifx \showISBNx    \undefined \def \showISBNx     #1{\unskip}     \fi
\ifx \showISBNxiii \undefined \def \showISBNxiii  #1{\unskip}     \fi
\ifx \showISSN     \undefined \def \showISSN      #1{\unskip}     \fi
\ifx \showLCCN     \undefined \def \showLCCN      #1{\unskip}     \fi
\ifx \shownote     \undefined \def \shownote      #1{#1}          \fi
\ifx \showarticletitle \undefined \def \showarticletitle #1{#1}   \fi
\ifx \showURL      \undefined \def \showURL       {\relax}        \fi
\providecommand\bibfield[2]{#2}
\providecommand\bibinfo[2]{#2}
\providecommand\natexlab[1]{#1}
\providecommand\showeprint[2][]{arXiv:#2}

\bibitem[\protect\citeauthoryear{Abramoff, Lavin, Birch, Shah, and
  Folk}{Abramoff et~al\mbox{.}}{2018}]%
        {abr:2018}
\bibfield{author}{\bibinfo{person}{M.D. Abramoff}, \bibinfo{person}{P.T.
  Lavin}, \bibinfo{person}{M. Birch}, \bibinfo{person}{N. Shah}, {and}
  \bibinfo{person}{J.C. Folk}.} \bibinfo{year}{2018}\natexlab{}.
\newblock \showarticletitle{Pivotal trial of an autonomous {AI}-based
  diagnostic system for detection of {D}iabetic {R}etinopathy in primary care
  offices}.
\newblock \bibinfo{journal}{\emph{Digital Medicine}} (\bibinfo{year}{2018}),
  \bibinfo{pages}{1--39}.
\newblock


\bibitem[\protect\citeauthoryear{Almotiri, Elleithy, and Elleithy}{Almotiri
  et~al\mbox{.}}{2018}]%
        {alm:2018}
\bibfield{author}{\bibinfo{person}{J. Almotiri}, \bibinfo{person}{K. Elleithy},
  {and} \bibinfo{person}{A. Elleithy}.} \bibinfo{year}{2018}\natexlab{}.
\newblock \showarticletitle{Retinal {V}essels {S}egmentation {T}echniques and
  {A}lgorithms: {A} {S}urvey}.
\newblock \bibinfo{journal}{\emph{Appl. Sci.}} \bibinfo{volume}{8},
  \bibinfo{number}{155} (\bibinfo{year}{2018}), \bibinfo{pages}{1--31}.
\newblock


\bibitem[\protect\citeauthoryear{Arnould, Guenancia, Alassane, Kawasaki, Daien,
  Tzourio, Kawasaki, Bourredjem, Bron, and Creuzot-Garcher}{Arnould
  et~al\mbox{.}}{2018}]%
        {arn:2018}
\bibfield{author}{\bibinfo{person}{C. Arnould, L.~Binquet}, \bibinfo{person}{C.
  Guenancia}, \bibinfo{person}{S. Alassane}, \bibinfo{person}{R. Kawasaki},
  \bibinfo{person}{V. Daien}, \bibinfo{person}{C. Tzourio}, \bibinfo{person}{Y.
  Kawasaki}, \bibinfo{person}{A. Bourredjem}, \bibinfo{person}{A. Bron}, {and}
  \bibinfo{person}{C. Creuzot-Garcher}.} \bibinfo{year}{2018}\natexlab{}.
\newblock \showarticletitle{Association between the retinal vascular network
  with {S}ingapore "{I}" {V}essel {A}ssessment ({S}{I}{V}{A}) software,
  cardiovascular history and risk factors in the elderly: The {M}ontrachet
  study, population-based study}.
\newblock \bibinfo{journal}{\emph{PLOS One}}  \bibinfo{volume}{3}
  (\bibinfo{date}{April} \bibinfo{year}{2018}), \bibinfo{pages}{1--14}.
\newblock


\bibitem[\protect\citeauthoryear{Brunner, Shipley, Witte, Fuller, and
  Marmot}{Brunner et~al\mbox{.}}{2006}]%
        {bru:2006}
\bibfield{author}{\bibinfo{person}{E. Brunner}, \bibinfo{person}{M. Shipley},
  \bibinfo{person}{D. Witte}, \bibinfo{person}{J. Fuller}, {and}
  \bibinfo{person}{M. Marmot}.} \bibinfo{year}{2006}\natexlab{}.
\newblock \showarticletitle{Relation between blood glucose and coronary
  mortality over 33 years in the {W}hitehall Study}.
\newblock \bibinfo{journal}{\emph{Diabetes Care}}  \bibinfo{volume}{29}
  (\bibinfo{year}{2006}), \bibinfo{pages}{26--31}.
\newblock


\bibitem[\protect\citeauthoryear{Cheung, Lamoureux, Ikram, Sasongko, Epi, Ding,
  Zheng, Mitchell, Wang, and Wong}{Cheung et~al\mbox{.}}{2012}]%
        {che:2012}
\bibfield{author}{\bibinfo{person}{C.Y. Cheung}, \bibinfo{person}{E.
  Lamoureux}, \bibinfo{person}{M.K. Ikram}, \bibinfo{person}{M.B. Sasongko},
  \bibinfo{person}{M. Epi}, \bibinfo{person}{J. Ding}, \bibinfo{person}{Y.
  Zheng}, \bibinfo{person}{P. Mitchell}, \bibinfo{person}{J.J. Wang}, {and}
  \bibinfo{person}{T.Y. Wong}.} \bibinfo{year}{2012}\natexlab{}.
\newblock \showarticletitle{Retinal {V}ascular {G}eometry in {A}sian {P}ersons
  with {D}iabetes and {R}etinopathy}.
\newblock \bibinfo{journal}{\emph{Journal of Diabetes Science and Technology}}
  \bibinfo{volume}{6}, \bibinfo{number}{3} (\bibinfo{year}{2012}).
\newblock


\bibitem[\protect\citeauthoryear{Cheung, Zheng, Hsu, Lee, Lau, Mitchell, Wang,
  Klein, and Wong}{Cheung et~al\mbox{.}}{2011}]%
        {cheung2011retinal}
\bibfield{author}{\bibinfo{person}{Carol Yim-lui Cheung},
  \bibinfo{person}{Yingfeng Zheng}, \bibinfo{person}{Wynne Hsu},
  \bibinfo{person}{Mong~Li Lee}, \bibinfo{person}{Qiangfeng~Peter Lau},
  \bibinfo{person}{Paul Mitchell}, \bibinfo{person}{Jie~Jin Wang},
  \bibinfo{person}{Ronald Klein}, {and} \bibinfo{person}{Tien~Yin Wong}.}
  \bibinfo{year}{2011}\natexlab{}.
\newblock \showarticletitle{Retinal vascular tortuosity, blood pressure, and
  cardiovascular risk factors}.
\newblock \bibinfo{journal}{\emph{Ophthalmology}} \bibinfo{volume}{118},
  \bibinfo{number}{5} (\bibinfo{year}{2011}), \bibinfo{pages}{812--818}.
\newblock


\bibitem[\protect\citeauthoryear{Cheung, Ong, Ikram, Ong, Lia, Hilal,
  S.Catindig, Venketasubramanian, Yap, Seow, Chen, and Wong}{Cheung
  et~al\mbox{.}}{2014}]%
        {che:2014}
\bibfield{author}{\bibinfo{person}{Y.L. Cheung}, \bibinfo{person}{Y.T. Ong},
  \bibinfo{person}{M.K. Ikram}, \bibinfo{person}{S.Y. Ong}, \bibinfo{person}{X.
  Lia}, \bibinfo{person}{S. Hilal}, \bibinfo{person}{J.S. S.Catindig},
  \bibinfo{person}{N. Venketasubramanian}, \bibinfo{person}{P. Yap},
  \bibinfo{person}{D. Seow}, \bibinfo{person}{C.P. Chen}, {and}
  \bibinfo{person}{T.Y. Wong}.} \bibinfo{year}{2014}\natexlab{}.
\newblock \showarticletitle{Microvascular network alterations in the retina of
  patients with {A}lzheimer's disease}.
\newblock \bibinfo{journal}{\emph{Alzheimer's \& Dementia}}
  \bibinfo{volume}{10} (\bibinfo{year}{2014}), \bibinfo{pages}{135--142}.
\newblock


\bibitem[\protect\citeauthoryear{Estrada, Allingham, Mettu, Cousins, Tomasi,
  and Farsiu}{Estrada et~al\mbox{.}}{2015}]%
        {estrada:2015}
\bibfield{author}{\bibinfo{person}{Rolando Estrada}, \bibinfo{person}{Michael~J
  Allingham}, \bibinfo{person}{Priyatham~S Mettu}, \bibinfo{person}{Scott~W
  Cousins}, \bibinfo{person}{Carlo Tomasi}, {and} \bibinfo{person}{Sina
  Farsiu}.} \bibinfo{year}{2015}\natexlab{}.
\newblock \showarticletitle{Retinal artery-vein classification via topology
  estimation}.
\newblock \bibinfo{journal}{\emph{IEEE transactions on medical imaging}}
  \bibinfo{volume}{34}, \bibinfo{number}{12} (\bibinfo{year}{2015}),
  \bibinfo{pages}{2518--2534}.
\newblock


\bibitem[\protect\citeauthoryear{Fraz, Remagnino, Hoppe, Uyyanonvara, Rudnicka,
  Owen, and Barman}{Fraz et~al\mbox{.}}{2012}]%
        {fraz:2012}
\bibfield{author}{\bibinfo{person}{M.M. Fraz}, \bibinfo{person}{P. Remagnino},
  \bibinfo{person}{A. Hoppe}, \bibinfo{person}{B. Uyyanonvara},
  \bibinfo{person}{A.R. Rudnicka}, \bibinfo{person}{C.G. Owen}, {and}
  \bibinfo{person}{S.A. Barman}.} \bibinfo{year}{2012}\natexlab{}.
\newblock \showarticletitle{Blood {V}essel {S}egmentation {M}ethodologies in
  {R}etinal {I}mages: {A} {S}urvey}.
\newblock \bibinfo{journal}{\emph{pubmed}} (\bibinfo{date}{Oct.}
  \bibinfo{year}{2012}).
\newblock


\bibitem[\protect\citeauthoryear{Frigo}{Frigo}{1999}]%
        {fri:1999}
\bibfield{author}{\bibinfo{person}{M. Frigo}.} \bibinfo{year}{1999}\natexlab{}.
\newblock \showarticletitle{A {F}ast {F}ourier {T}ransform {C}ompiler}.
\newblock \bibinfo{journal}{\emph{Proceedings of the 1999 ACM SIGPLAN
  Conference on Programming Language Design and Implementation}}
  (\bibinfo{year}{1999}).
\newblock


\bibitem[\protect\citeauthoryear{Gulshan, Peng, Coram, Stumpe, Wu,
  Narayanaswamy, Venugopalan, Widner, Madams, and Cuadros}{Gulshan
  et~al\mbox{.}}{2016}]%
        {gul:2016}
\bibfield{author}{\bibinfo{person}{V. Gulshan}, \bibinfo{person}{L. Peng},
  \bibinfo{person}{M. Coram}, \bibinfo{person}{M.C. Stumpe},
  \bibinfo{person}{D. Wu}, \bibinfo{person}{A. Narayanaswamy},
  \bibinfo{person}{S. Venugopalan}, \bibinfo{person}{K. Widner},
  \bibinfo{person}{T. Madams}, {and} \bibinfo{person}{J. Cuadros}.}
  \bibinfo{year}{2016}\natexlab{}.
\newblock \showarticletitle{Development and validation of a {D}eep {L}earning
  algorithm for detection of {D}iabetic {R}etinopathy in retinal fundus
  photographs}.
\newblock \bibinfo{journal}{\emph{JAMA}} \bibinfo{volume}{316},
  \bibinfo{number}{22} (\bibinfo{year}{2016}), \bibinfo{pages}{2402--2410}.
\newblock


\bibitem[\protect\citeauthoryear{Hanff, Sharrett, Mosley, Shibata, Knopman,
  Klein, Klein, and Gottesman}{Hanff et~al\mbox{.}}{2014}]%
        {han:2014}
\bibfield{author}{\bibinfo{person}{T.C. Hanff}, \bibinfo{person}{A.R.
  Sharrett}, \bibinfo{person}{T.H. Mosley}, \bibinfo{person}{D. Shibata},
  \bibinfo{person}{D.S. Knopman}, \bibinfo{person}{R. Klein},
  \bibinfo{person}{B.E.K. Klein}, {and} \bibinfo{person}{R.F Gottesman}.}
  \bibinfo{year}{2014}\natexlab{}.
\newblock \showarticletitle{{R}etinal {M}icrovascular {A}bnormalities {P}redict
  {P}rogression of {B}rain {M}icrovascular {D}isease: {A}n {ARIC} {MRI}
  {S}tudy}.
\newblock \bibinfo{journal}{\emph{Stroke}} \bibinfo{volume}{45},
  \bibinfo{number}{4} (\bibinfo{date}{April} \bibinfo{year}{2014}),
  \bibinfo{pages}{1012--1017}.
\newblock


\bibitem[\protect\citeauthoryear{He, Zhang, Ren, and Sun}{He
  et~al\mbox{.}}{2015}]%
        {he:2015}
\bibfield{author}{\bibinfo{person}{K. He}, \bibinfo{person}{X. Zhang},
  \bibinfo{person}{S. Ren}, {and} \bibinfo{person}{J. Sun}.}
  \bibinfo{year}{2015}\natexlab{}.
\newblock \showarticletitle{Deep {R}esidual {L}earning for {I}mage
  {R}ecognition}.
\newblock \bibinfo{journal}{\emph{CoRR}}  \bibinfo{volume}{abs/1512.03385}
  (\bibinfo{year}{2015}).
\newblock
\showeprint[arxiv]{1512.03385}
\urldef\tempurl%
\url{http://arxiv.org/abs/1512.03385}
\showURL{%
\tempurl}


\bibitem[\protect\citeauthoryear{Ho, Cheung, Sabanayagam, Yip, Ikram, Ong,
  Mitchell, Chow, Cheng, Tai, and Wong}{Ho et~al\mbox{.}}{2017}]%
        {ho:2017}
\bibfield{author}{\bibinfo{person}{H. Ho}, \bibinfo{person}{C.Y Cheung},
  \bibinfo{person}{C. Sabanayagam}, \bibinfo{person}{W. Yip},
  \bibinfo{person}{M.K. Ikram}, \bibinfo{person}{P.G. Ong}, \bibinfo{person}{P.
  Mitchell}, \bibinfo{person}{K.Y. Chow}, \bibinfo{person}{C.Y. Cheng},
  \bibinfo{person}{E.S. Tai}, {and} \bibinfo{person}{T.Y. Wong}.}
  \bibinfo{year}{2017}\natexlab{}.
\newblock \showarticletitle{Retinopathy signs improved prediction and
  reclassification of cardiovascular disease risk in diabetes: A prospective
  cohort study}.
\newblock \bibinfo{journal}{\emph{Scientific Reports}} \bibinfo{volume}{7},
  \bibinfo{number}{41492} (\bibinfo{year}{2017}).
\newblock


\bibitem[\protect\citeauthoryear{Hsia, Margolis, Eaton, Wenger, Allison, Wu,
  and Black}{Hsia et~al\mbox{.}}{2007}]%
        {hsi:2007}
\bibfield{author}{\bibinfo{person}{J. Hsia}, \bibinfo{person}{K. Margolis},
  \bibinfo{person}{C. Eaton}, \bibinfo{person}{N. Wenger}, \bibinfo{person}{M.
  Allison}, \bibinfo{person}{L. Wu}, {and} \bibinfo{person}{H. Black}.}
  \bibinfo{year}{2007}\natexlab{}.
\newblock \showarticletitle{Prehypertension and cardiovascular disease risk in
  the Women's Health Initiative}.
\newblock \bibinfo{journal}{\emph{Circulation}}  \bibinfo{volume}{115}
  (\bibinfo{year}{2007}), \bibinfo{pages}{855--860}.
\newblock


\bibitem[\protect\citeauthoryear{Ikram, de~Jong, Vingerling, Witteman, Hofman,
  and Breteler}{Ikram et~al\mbox{.}}{2004}]%
        {ikr:2004}
\bibfield{author}{\bibinfo{person}{M.K. Ikram}, \bibinfo{person}{F.J. de Jong},
  \bibinfo{person}{J.R. Vingerling}, \bibinfo{person}{J.C. Witteman},
  \bibinfo{person}{A. Hofman}, {and} \bibinfo{person}{M.M. Breteler}.}
  \bibinfo{year}{2004}\natexlab{}.
\newblock \showarticletitle{Are retinal arteriolar or venular diameters
  associated with markers for cardiovascular disorders? The Rotterdam Study}.
\newblock \bibinfo{journal}{\emph{Invest Ophthalmol Vis Sci}}
  \bibinfo{volume}{45} (\bibinfo{year}{2004}), \bibinfo{pages}{2129--2134}.
\newblock


\bibitem[\protect\citeauthoryear{Ikram, Cheung, Lorenzi, Klein, Jones, and
  Wong}{Ikram et~al\mbox{.}}{2017}]%
        {ikr:2017}
\bibfield{author}{\bibinfo{person}{N.K. Ikram}, \bibinfo{person}{C.Y. Cheung},
  \bibinfo{person}{M. Lorenzi}, \bibinfo{person}{R. Klein},
  \bibinfo{person}{T.L.Z Jones}, {and} \bibinfo{person}{T.Y. Wong}.}
  \bibinfo{year}{2017}\natexlab{}.
\newblock \showarticletitle{Retinal vascular caliber as a biomarker for
  diabetes microvascular complications}.
\newblock \bibinfo{journal}{\emph{Diabetes Care}}  \bibinfo{volume}{36}
  (\bibinfo{year}{2017}), \bibinfo{pages}{750--759}.
\newblock


\bibitem[\protect\citeauthoryear{Jenkins, Joglekar, Hardikar, Keech, O'Neal,
  and Januszewski}{Jenkins et~al\mbox{.}}{2015}]%
        {jen:2015}
\bibfield{author}{\bibinfo{person}{A.J. Jenkins}, \bibinfo{person}{M.V.
  Joglekar}, \bibinfo{person}{A.A. Hardikar}, \bibinfo{person}{A.C. Keech},
  \bibinfo{person}{D.N O'Neal}, {and} \bibinfo{person}{A.S. Januszewski}.}
  \bibinfo{year}{2015}\natexlab{}.
\newblock \showarticletitle{Biomarkers in {D}iabetic {R}etinopathy}.
\newblock \bibinfo{journal}{\emph{The Review of Diabetic Studies}}
  \bibinfo{volume}{12} (\bibinfo{year}{2015}), \bibinfo{pages}{159--195}.
\newblock


\bibitem[\protect\citeauthoryear{Knudtson, Lee, Hubbard, Wong, Klein, and
  Klein}{Knudtson et~al\mbox{.}}{2003}]%
        {knu:2003}
\bibfield{author}{\bibinfo{person}{M.D. Knudtson}, \bibinfo{person}{K.E. Lee},
  \bibinfo{person}{L.D. Hubbard}, \bibinfo{person}{T.Y. Wong},
  \bibinfo{person}{R. Klein}, {and} \bibinfo{person}{B.E. Klein}.}
  \bibinfo{year}{2003}\natexlab{}.
\newblock \showarticletitle{Revised formulas for summarizing retinal vessel
  diameters}.
\newblock \bibinfo{journal}{\emph{Curr Eye Res.}}  \bibinfo{volume}{27}
  (\bibinfo{year}{2003}), \bibinfo{pages}{143--149}.
\newblock


\bibitem[\protect\citeauthoryear{Lim, Chee, Cheung, and Wong}{Lim
  et~al\mbox{.}}{2017}]%
        {lim:2017}
\bibfield{author}{\bibinfo{person}{L.S. Lim}, \bibinfo{person}{M.L. Chee},
  \bibinfo{person}{C.Y. Cheung}, {and} \bibinfo{person}{T.Y. Wong}.}
  \bibinfo{year}{2017}\natexlab{}.
\newblock \showarticletitle{Retinal {V}essel {G}eometry and the {I}ncidence and
  {P}rogression of {D}iabetic {R}etinopathy}.
\newblock \bibinfo{journal}{\emph{Investigative Ophthalmology \& Visual
  Science}} \bibinfo{volume}{58}, \bibinfo{number}{6} (\bibinfo{year}{2017}),
  \bibinfo{pages}{200--205}.
\newblock


\bibitem[\protect\citeauthoryear{Lin, Blumencranz, Brothers, and Grosvenor}{Lin
  et~al\mbox{.}}{2002}]%
        {lin:2002}
\bibfield{author}{\bibinfo{person}{D.Y. Lin}, \bibinfo{person}{M.S.
  Blumencranz}, \bibinfo{person}{R.J. Brothers}, {and} \bibinfo{person}{D.M.
  Grosvenor}.} \bibinfo{year}{2002}\natexlab{}.
\newblock \showarticletitle{The {S}ensitivity and {S}pecificity of
  {S}ingle-field {N}onmydriatic {M}onochromatic {D}igital {F}undus
  {P}hotography {W}ith {R}emote {I}mage {I}nterpretation for {D}iabetic
  {R}etinopathy {S}creening: {A} {C}omparison {W}ith {O}phthalmoscopy and
  {S}tandardized {M}ydriatic {C}olor {P}hotography}.
\newblock \bibinfo{journal}{\emph{Am J Ophthalmol.}} \bibinfo{volume}{134},
  \bibinfo{number}{2} (\bibinfo{year}{2002}), \bibinfo{pages}{204--213}.
\newblock


\bibitem[\protect\citeauthoryear{MacGillivray, Trucco, Cameron, Dhillon,
  Houston, and Van~Beek}{MacGillivray et~al\mbox{.}}{2014}]%
        {macgillivray:2014}
\bibfield{author}{\bibinfo{person}{TJ MacGillivray}, \bibinfo{person}{Emanuele
  Trucco}, \bibinfo{person}{JR Cameron}, \bibinfo{person}{Baljean Dhillon},
  \bibinfo{person}{JG Houston}, {and} \bibinfo{person}{EJR Van~Beek}.}
  \bibinfo{year}{2014}\natexlab{}.
\newblock \showarticletitle{Retinal imaging as a source of biomarkers for
  diagnosis, characterization and prognosis of chronic illness or long-term
  conditions}.
\newblock \bibinfo{journal}{\emph{The British journal of radiology}}
  \bibinfo{volume}{87}, \bibinfo{number}{1040} (\bibinfo{year}{2014}),
  \bibinfo{pages}{20130832}.
\newblock


\bibitem[\protect\citeauthoryear{Mansburger, Sheppler, Barker, Gardiner,
  Demirel, Wooten, and Becker}{Mansburger et~al\mbox{.}}{2015}]%
        {man:2015}
\bibfield{author}{\bibinfo{person}{S.L. Mansburger}, \bibinfo{person}{C.
  Sheppler}, \bibinfo{person}{S.K. Barker}, \bibinfo{person}{S.K. Gardiner},
  \bibinfo{person}{S. Demirel}, \bibinfo{person}{K. Wooten}, {and}
  \bibinfo{person}{T. Becker}.} \bibinfo{year}{2015}\natexlab{}.
\newblock \showarticletitle{{L}ong-term {C}omparative {E}ffectiveness of
  {T}elemedicine in {P}roviding {D}iabetic {R}etinopathy {S}creening
  {E}xaminations {A} {R}andomized {C}linical {T}rial}.
\newblock \bibinfo{journal}{\emph{JAMA Ophthalmol.}} \bibinfo{volume}{133},
  \bibinfo{number}{5} (\bibinfo{year}{2015}), \bibinfo{pages}{518--525}.
\newblock


\bibitem[\protect\citeauthoryear{Reports}{Reports}{2017}]%
        {nvs:2017}
\bibfield{author}{\bibinfo{person}{National Vital~Statistics Reports}.}
  \bibinfo{year}{2017}\natexlab{}.
\newblock \bibinfo{booktitle}{}.
\newblock \bibinfo{type}{{T}echnical {R}eport}~66.
\newblock
\urldef\tempurl%
\url{https://www.cdc.gov/nchs/data/nvsr/nvsr66/nvsr66_06.pdf}
\showURL{%
\tempurl}


\bibitem[\protect\citeauthoryear{Ronneberger, Fischer, and Brox}{Ronneberger
  et~al\mbox{.}}{2015}]%
        {olaf:2015}
\bibfield{author}{\bibinfo{person}{O. Ronneberger}, \bibinfo{person}{P.
  Fischer}, {and} \bibinfo{person}{T. Brox}.} \bibinfo{year}{2015}\natexlab{}.
\newblock \showarticletitle{U-{N}et: {C}onvolutional {N}etworks for
  {B}iomedical {I}mage {S}egmentation}.
\newblock \bibinfo{journal}{\emph{arXiv}} (\bibinfo{date}{May}
  \bibinfo{year}{2015}).
\newblock


\bibitem[\protect\citeauthoryear{Sayres, Taly, Rahimy, Blumer, Coz, Hammel,
  Krause, and Narayanaswamy}{Sayres et~al\mbox{.}}{2019}]%
        {say:2019}
\bibfield{author}{\bibinfo{person}{R. Sayres}, \bibinfo{person}{A. Taly},
  \bibinfo{person}{E. Rahimy}, \bibinfo{person}{K. Blumer}, \bibinfo{person}{D.
  Coz}, \bibinfo{person}{N. Hammel}, \bibinfo{person}{J. Krause}, {and}
  \bibinfo{person}{A. Narayanaswamy}.} \bibinfo{year}{2019}\natexlab{}.
\newblock \showarticletitle{Using a {D}eep {L}earning algorithm and integrated
  gradients explanation to assist grading for {D}iabetic {R}etinophathy}.
\newblock \bibinfo{journal}{\emph{Ophthalmology}} \bibinfo{volume}{126},
  \bibinfo{number}{4} (\bibinfo{date}{April} \bibinfo{year}{2019}),
  \bibinfo{pages}{552--564}.
\newblock


\bibitem[\protect\citeauthoryear{Scanlon}{Scanlon}{2017}]%
        {scan:2017}
\bibfield{author}{\bibinfo{person}{P.H. Scanlon}.}
  \bibinfo{year}{2017}\natexlab{}.
\newblock \showarticletitle{The {E}nglish {N}ational {S}creening {P}rogramme
  for {D}iabetic {R}etinopathy}.
\newblock \bibinfo{journal}{\emph{Acta Diabetol}}  \bibinfo{volume}{54}
  (\bibinfo{year}{2017}), \bibinfo{pages}{515--525}.
\newblock


\bibitem[\protect\citeauthoryear{Soomro, Afifi, Zheng, Soomro, Gao, Hellwich,
  and Paul}{Soomro et~al\mbox{.}}{2019}]%
        {soo:2019}
\bibfield{author}{\bibinfo{person}{T.A. Soomro}, \bibinfo{person}{A.J. Afifi},
  \bibinfo{person}{L. Zheng}, \bibinfo{person}{S. Soomro}, \bibinfo{person}{J.
  Gao}, \bibinfo{person}{O. Hellwich}, {and} \bibinfo{person}{M. Paul}.}
  \bibinfo{year}{2019}\natexlab{}.
\newblock \showarticletitle{Deep {L}earning {M}odels for {R}etinal {B}lood
  {V}essels {S}egmentation: {A} {R}eview}.
\newblock \bibinfo{journal}{\emph{IEEE Access}}  \bibinfo{volume}{7}
  (\bibinfo{year}{2019}), \bibinfo{pages}{71696--71717}.
\newblock


\bibitem[\protect\citeauthoryear{Thomas, Dunstan, Luzio, Chowdhury, North,
  Hale, Gibbins, and Owens}{Thomas et~al\mbox{.}}{2015}]%
        {tho:2015}
\bibfield{author}{\bibinfo{person}{R.L. Thomas}, \bibinfo{person}{F.D.
  Dunstan}, \bibinfo{person}{S.D. Luzio}, \bibinfo{person}{S.R. Chowdhury},
  \bibinfo{person}{R.V. North}, \bibinfo{person}{S.L. Hale},
  \bibinfo{person}{R.L. Gibbins}, {and} \bibinfo{person}{D.R. Owens}.}
  \bibinfo{year}{2015}\natexlab{}.
\newblock \showarticletitle{Prevalence of {D}iabetic {R}etinopathy within a
  national {D}iabetic {R}etinopathy screening service}.
\newblock \bibinfo{journal}{\emph{Br J Ophthalmology}} \bibinfo{volume}{99},
  \bibinfo{number}{1} (\bibinfo{year}{2015}), \bibinfo{pages}{64--68}.
\newblock


\bibitem[\protect\citeauthoryear{Wong, Kamineni, Klein, Sharrett, Klein,
  Siscovick, Cushman, and Duncan}{Wong et~al\mbox{.}}{2017}]%
        {won:2017}
\bibfield{author}{\bibinfo{person}{T.Y. Wong}, \bibinfo{person}{A. Kamineni},
  \bibinfo{person}{R. Klein}, \bibinfo{person}{A.R. Sharrett},
  \bibinfo{person}{B.E. Klein}, \bibinfo{person}{D.S. Siscovick},
  \bibinfo{person}{M. Cushman}, {and} \bibinfo{person}{B.B. Duncan}.}
  \bibinfo{year}{2017}\natexlab{}.
\newblock \showarticletitle{Quantitative {R}etinal {V}enular {C}aliber and
  {R}isk of {C}ardiovascular {D}isease in {O}lder {P}ersons}.
\newblock \bibinfo{journal}{\emph{Arch Int Med.}} \bibinfo{volume}{166},
  \bibinfo{number}{21} (\bibinfo{year}{2017}), \bibinfo{pages}{2388--2394}.
\newblock


\bibitem[\protect\citeauthoryear{Zahn and Roskies}{Zahn and Roskies}{1972}]%
        {zah:1972}
\bibfield{author}{\bibinfo{person}{C.T. Zahn} {and} \bibinfo{person}{R.Z.
  Roskies}.} \bibinfo{year}{1972}\natexlab{}.
\newblock \showarticletitle{Fourier descriptors for plane closed curves}.
\newblock \bibinfo{journal}{\emph{IEEE Trans. Comput.}} \bibinfo{volume}{100},
  \bibinfo{number}{3} (\bibinfo{date}{March} \bibinfo{year}{1972}),
  \bibinfo{pages}{269--281}.
\newblock


\bibitem[\protect\citeauthoryear{Zeiler and Fergus}{Zeiler and Fergus}{2013}]%
        {zei:2013}
\bibfield{author}{\bibinfo{person}{M.D. Zeiler} {and} \bibinfo{person}{R.
  Fergus}.} \bibinfo{year}{2013}\natexlab{}.
\newblock \showarticletitle{Visualizing and {U}nderstanding {C}onvolutional
  {N}etworks}.
\newblock \bibinfo{journal}{\emph{CoRR}}  \bibinfo{volume}{abs/1311.2901}
  (\bibinfo{year}{2013}).
\newblock
\showeprint[arxiv]{1311.2901}
\urldef\tempurl%
\url{http://arxiv.org/abs/1311.2901}
\showURL{%
\tempurl}


\end{thebibliography}
\appendix
\section{Appendix I - Harmonic Descriptors}
\footnotesize
\label{theory}
Consider a periodic signal $z(t)$ with period $T$. The period being $T$, we will take the fundamental frequency to be $\omega_0=\frac{2\pi}{T}$. We may represent any such function (with some very minor restrictions) using a Fourier series. There are two common representations of Fourier series, ``Trigonometric'' and ``Exponential''. For easy reference the two forms are
\begin{equation}
\begin{aligned}
z(t) &=a_0 + \sum_{k=1}^\infty a_k\cos(k\omega_0t) + b_k\sin(k\omega_0t) &&\text{\it{(Trigonometric)}}  \\
&=\sum_{k=-\infty}^\infty c_k e^{i k\omega_0t} &&\text{\it{(Exponential)}}.
\end{aligned}
\end{equation}
The coefficients $a_k, b_k$ and $c_k$ are related by the following equations\footnote{$*$ is the complex conjugate operator.}
\begin{gather}
c_0=a_0,\label{eq:c0}\\
c_k = \frac{a_k}{2} - i~\frac{b_k}{2},\text{ and } c_{-k}=c_k^*\label{eq:ck}
\end{gather}
for $k\ne0$ with $i=\sqrt{-1}$, and $a_0=\frac{1}{T}\int_T y(t)dt$, $a_k=\frac{2}{T}\int_T y(t)\cos(k\omega_0t)dt$, and $b_k=\frac{2}{T}\int_T y(t)\sin(k\omega_0t)dt$. A leaner representation of the Fourier Series using complex exponential ends up with the following
\begin{equation*}
\label{eq:complex}
z(t)=\sum_{k=-\infty}^\infty c_k e^{i k\omega_0t}\text{ where }c_k=\frac{1}{T}\int_T y(t)e^{i k\omega_0t}dt.\\
\end{equation*}
\subsection{Vessel Transformation}
Let $\mathcal{C}=\left\{(x(t),y(t))\mid t=0,\ldots,N\right\}\subset\mathbb{R}^2$ be a discrete curve that can be reduced to a discrete function $\phi:[0,N]\to\mathbb{R}$ after a rotation $\rho$ and a translation $\tau$. Its coordinates can be then viewed as the sampling values
\begin{equation*}
(x(t),y(t))=(\tau^{-1}\circ\rho^{-1})\left[f\left(\frac{\pi}{N}(2t-N)\right)\right]
\end{equation*}
of a continuous curve $f:[-\pi,\pi]\to\mathbb{R}$ such that $f$ can be extended continuously to a $2\pi$-periodic function. When the function $z$ is expanded into a Fourier series, a fixed number of discrete Fourier coefficients approximately represents the curve. Hence, among other interesting properties, allowing for data reduction. For the rest of the paper let us assume that we are dealing with periodic smooth functions $\mathbb{R}\to\mathbb{R}$ having $[-\pi,\pi]$ as their fundamental period $T$.
\subsection{Harmonic Descriptors}
When identifying the two-dimensional plane $\mathbb{R}^2$ with the complex plane $\mathbb{C}$, a curve may be approximated by a sequence of complex numbers $z(t)=x(t)+i\cdot y(t)$ having the discrete Fourier expansion $t=0,\ldots,N-1$:
\begin{equation*}
z_N(t)=\sum_{k=0}^{N-1}\hat{c}_ke^{\frac{2\pi}{N}ikt}
\end{equation*}
with discrete Fourier coefficients
\begin{equation*}
\hat{c}_k=\frac{1}{N}\sum_{t=0}^{N-1}z(t)e^{-\frac{2\pi}{N}ikt}.
\end{equation*}
When the coefficients $\hat{c}_k$ are interpreted as numerical approximations of the Fourier coefficients $c_k$ of the continuous curve $z(t)=x(\frac{N}{2\pi}t) +i\cdot y(\frac{N}{2\pi}t)$ 
\begin{equation*}
c_k=\frac{1}{2\pi}\int_{-\pi}^{\pi} z(t)e^{i kt}dt
\end{equation*}
then the connection between $\hat{c}_k$ and $c_k$ is given by
\begin{equation*}
\label{eq:approximation}
\begin{aligned}
\hat{c}_k&\approx c_k&\text{ for }0\leq k<\frac{N}{2},\\
\hat{c}_{N-k}&\approx c_{-k}&\text{ for }1\leq k<\frac{N}{2}.
\end{aligned}
\end{equation*}

\subsection{Derivative's Energy}
\label{sec:energy}
It is known from Parseval's identity that the formula $\mathcal{E}_z=\frac{1}{2\pi}\int_{-\pi}^\pi|z(t)|^2dt=\sum_{k=-\infty}^\infty|c_k|^2$ represents the total \textit{energy} of the waveform $z(t)$ over $[-\pi,\pi]$. In the above, all Harmonic Descriptors have the same contribution to the total, namely 1. Since $|c'_k|^2=k^2|c_k|^2\implies\mathcal{E}'=\frac{1}{2\pi}\int_{-\pi}^{\pi}z'(t)^2dt=\sum_{k=-\infty}^\infty k^2|c_k|^2$ we derive instead a weighted version of the energy $\mathcal{E}$, noted $\mathcal{E}'_{z_N}$, by the formula
\begin{equation}
\label{eq:weighted-energy}
\mathcal{E}'_{z_N}\stackrel{\text{\tiny def}}{=}d_{z_N}^2\sum_{k=-N}^N k^2|c_k|^2=\frac{d_{z_N}^2}{2}\sum_{k=0}^N k^2(a_k^2+b_k^2)
\end{equation}
where $d_z$ is the scaling factor used for transforming the vessel $z(t)$ over $[-\pi,\pi]$. Notice that the DC component $c_0$ makes no contribution whatsoever to the value of $\mathcal{E}'_{z_N}$, obviously a corollary of the HD translation invariance principle. If $t\mapsto z'(t)^2$ is integrable over $[-\pi,\pi]$ the infinite series $\sum_{k=-\infty}^\infty k^2|c_k|^2$ converges since it's then a positive increasing series bounded  by $\frac{1}{\pi}\int_{-\pi}^{\pi}z'(t)^2dt<\infty$.

\subsection{Tortuosity}
\label{sec:tortuosity}
Tortuosity is the result of an accumulation of curvature along a blood vessel length and vessel tortuosity augmentation is a first class manifestation of the changes in vessel morphology. It could be caused by several diseases like high blood pressure, angiogenesis, or blood vessel congestion.
\subsubsection{Definition}
\begin{definition}[Tortuosity]
	\label{def:tortuosity}
	Let $z_N(t)$ defined on $[-\pi,\pi]$. The tortuosity $\tau_{z_N}$ of $z_N(t)$ is defined as 
	\begin{equation}
	\label{eq:tortuosity}
	\tau_{z_N}\stackrel{\text{\tiny def}}{=}\mathcal{E}'_{z_N}=d_{z_N}^2\sum_{k=-N}^N k^2|c_k|^2.
	\end{equation}
\end{definition}

Tortuosity definition involves also second derivatives since 
\begin{equation}
\label{eq:second-derivative}
\sum_{k=-N}^N k^2|c_k|^2= \sum_{\stackrel{k=-N}{k\neq0}}^N\frac{1}{k^2}|c''_k|^2.
\end{equation}
In \eqref{eq:second-derivative}, notice how low-frequencies matter more than high-frequencies.

It can be said that tortuosity is a global Fourier expression of the curvature of a vessel. For the sake of the reader's curiosity, Table~\ref{table:tortuosity} gives a list of the tortuosity value for quite a few simple functions.

As the spatial domain contents rotates the frequency domain contents we may deduce that tortuosity is rotation invariant. Geometrically, rotation in the frequency domain is equivalent to multiply each HD $c_k$ by a common factor $e^{i\theta}$, where $\theta$ is the specified angle of rotation. Hence $\sum_{k=-N}^N k^2|e^{i\theta}c_k|^2=|e^{i\theta}|^2\sum_{k=-N}^N k^2|c_k|^2=\sum_{k=-N}^N k^2|c_k|^2$ since $|e^{i\theta}|=1$. In summary,
\subsubsection{Properties}
\begin{itemize}
	\item \textbf{Translation}: Tortuosity is translation invariant.
	\item \textbf{Rotation}: Tortuosity is rotation invariant.
	\item \textbf{Scaling}: Tortuosity is a quadratic function of the scaling factor.
\end{itemize}

\subsection{Tortuosity of Some Common Curves}
\begin{table}[!h]
	\caption{\small Tortuosity of Some Common Curves Sharing the Same Amplitude ($=2$) for $t\in[-\pi,\pi]$}
	\label{table:tortuosity}
	\footnotesize
	\begin{tabular}{c|c|c|c} 
		\textbf{Curve Type}&\textbf{Definition}&\textbf{Tortuosity}&\textbf{Value}\\
		\hline
		\textit{Trigonometric}&$\sin(jt), \cos(jt)$&$\frac{j^2}{2}$&2 ($j=2)$\\
		\textit{Inverse Trig.}&$\frac{\arctan(t)}{\arctan(\pi)}$&$\frac{\left({\pi}^2+1\right)\arctan\left({\pi}\right)+{\pi}}{{\pi}\left(2{\pi}^2+2\right)\arctan\left({\pi}\right)}$&$0.195\ldots$\\
		\textit{Paraboloid}&$\frac{2}{\pi^2}(\pi^2-t^2)$&$\frac{8}{3}$&$2.666\ldots$\\
		\textit{Hyperboloid}&$2\frac{\sqrt{1+t^2}}{\sqrt{1+\pi^2}-1}$&$-\frac{4\left(\arctan\left({\pi}\right)-{\pi}\right)}{\sqrt{{\pi}^2+1}-1}$&$3.272\ldots$\\
		\textit{Elliptical Arc\footnote{Here $a=-\pi+\epsilon$ and $b=\pi-\epsilon$ with $\epsilon=0.999$.}}&$\alpha_2(t)$&$4\cdot\frac{\ln(2\times10^3)-2}{\pi^2}$&$2.270\ldots$\\
		\textit{Gaussian}&$2e^{-\frac{t^2}{2}}$&$2\sqrt{2\pi}\sum\limits_{k=0}^\infty k^2e^{-k^2}$&$2.029\ldots$\\
	\end{tabular}
\end{table}
See Table~\ref{table:tortuosity}. The inverse trigonometric function $\arctan(t)$ almost being a straight line on $[-\pi,\pi]$ it's natural that its tortuosity value is near zero.

\subsection{Tortuosity of the Union of Two Vessels}
\begin{theorem}[Tortuosity of  the Union of Two Vessels]
	\label{ltheorem:union}
	Let $Z=U\cup V$, $d_z$, $d_u$, and $d_v$ defined as above. The tortuosity $\tau_z$ of the union $Z$ of $U$ and $V$ is equal to
	\begin{equation}
	\label{eq:union}
	\begin{aligned}
	\tau_z&=d_z^2\left[\frac{\tau_u}{d_u^2}+\frac{\tau_v}{d_v^2}\right].\\
	\end{aligned}
	\end{equation}
\end{theorem}
As a special case, if $d_z=d_u=d_v$ then \eqref{eq:union} reduces to
\begin{equation}
\tau_z=\tau_u+\tau_v.
\end{equation}
\FloatBarrier
\section[ROCS]{Appendix II - ROC Curves}
\begin{figure}[!h]
	\centering
	\includegraphics[width=\columnwidth]{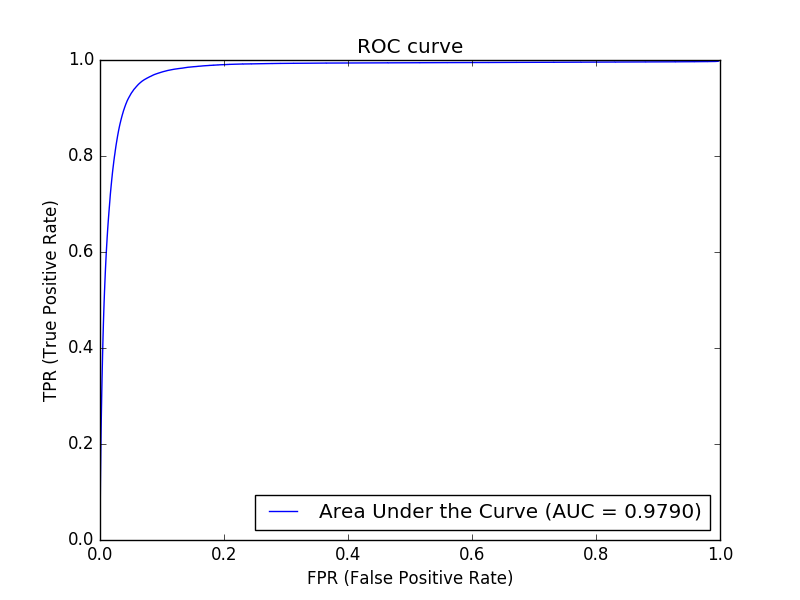}%
	\caption{\small ROC curve for semantic segmentation of 150 training images}
	\label{fig:semantic_roc}
\end{figure}
\FloatBarrier
\begin{figure}
	\centering
	\begin{subfigure}{\columnwidth}
	\centering
	\includegraphics[width=\columnwidth]{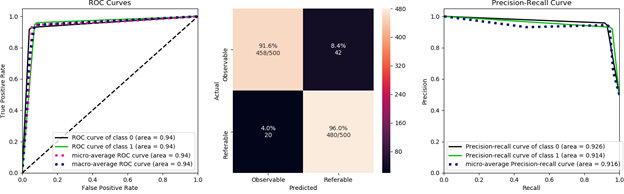}%
	\caption{I-A ROC Curve}
	\end{subfigure}
	\begin{subfigure}{\columnwidth}
	\centering
	\includegraphics[width=\columnwidth]{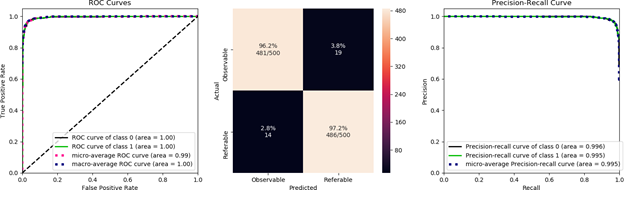}%
	\caption{I-B ROC Curve}
	\end{subfigure}
	\caption{\small ROC curves for scenario I}
	\label{fig:I_ROCS}
\end{figure}
\FloatBarrier
\begin{figure}
	\centering
	\begin{subfigure}{.45\textwidth}
	\centering
	\includegraphics[width=\columnwidth]{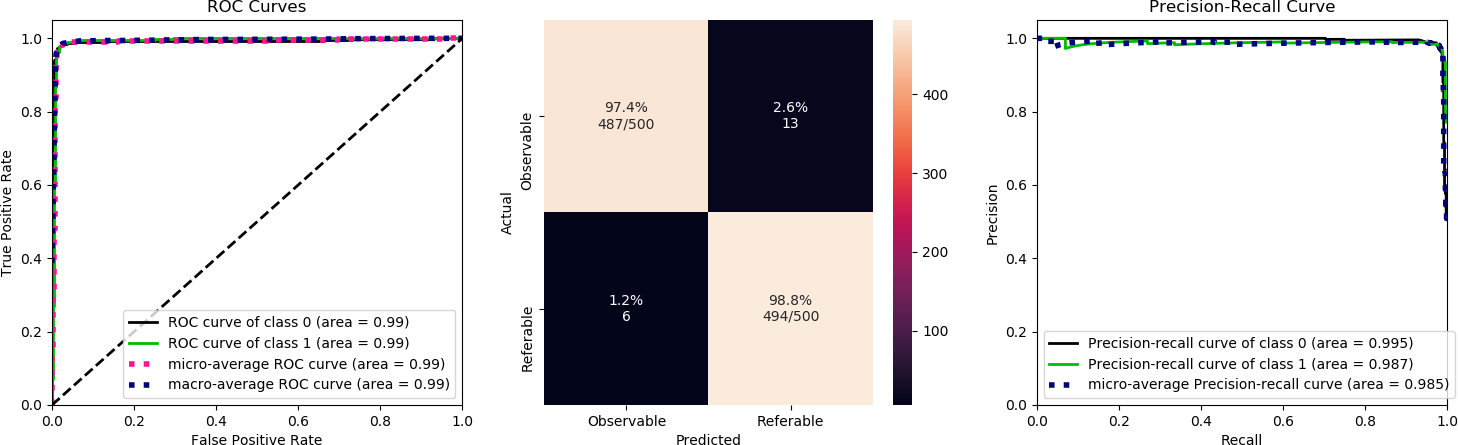}%
	\caption{II-A ROC Curve}
	\end{subfigure}
	\begin{subfigure}{.45\textwidth}
	\centering
	\includegraphics[width=\columnwidth]{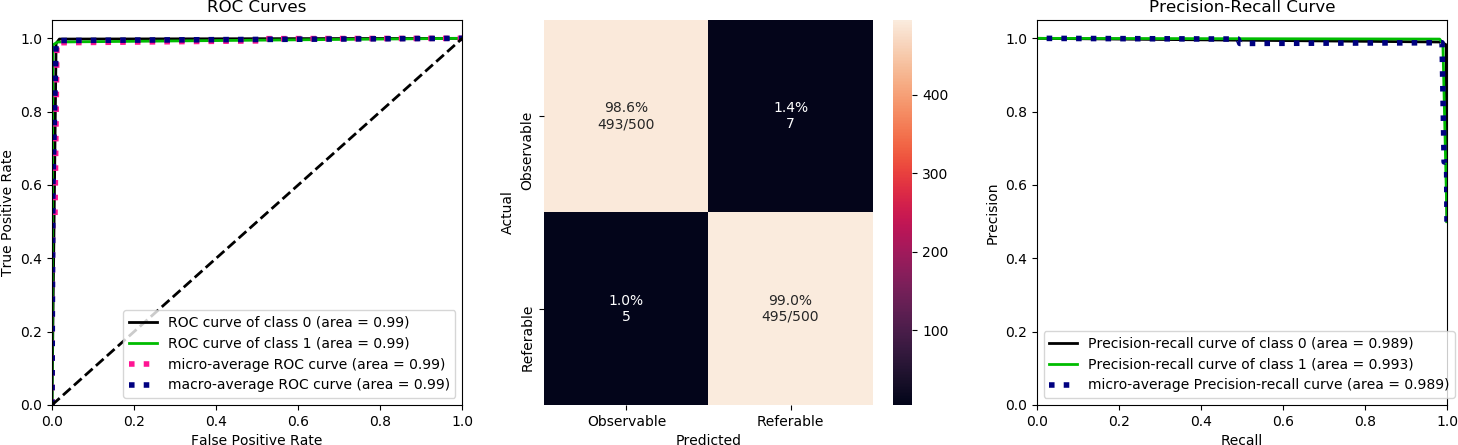}%
	\caption{II-B ROC Curve}
	\end{subfigure}
	\caption{\small ROC curves for scenario II}
	\label{fig:II_ROCS}
\end{figure}
\end{document}